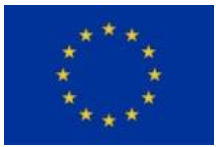 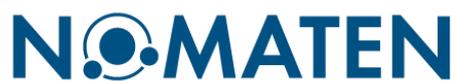

This work was carried out in whole or in part within the framework of the NOMATEN Centre of Excellence, supported from the European Union Horizon 2020 research and innovation program (Grant Agreement No. 857470) and from the European Regional Development Fund via the Foundation for Polish Science International Research Agenda PLUS program (Grant No. MAB PLUS/2018/8), and the Ministry of Science and Higher Education's initiative "Support for the Activities of Centers of Excellence Established in Poland under the Horizon 2020 Program" (agreement no. MEiN/2023/DIR/3795).

The version of record of this article, first published in Diamond and Related Materials, Volume 146, June 2024, 111247, is available online at Publisher's website: https://dx.doi.org/10.1016/j.diamond.2024.111247

# Evolution of radiation-induced damage in nuclear graphite – a comparative structural and microstructural study

Magdalena Wilczopolska[1,2], Kinga Suchorab[1,2], Magdalena Gawęda[1*], Małgorzata Frelek-Kozak[1,2], Paweł Ciepielewski[3], Marcin Brykała[2], Wojciech Chmurzyński[2], Iwona Jóźwik[1]

[1]*NOMATEN CoE, NOMATEN MAB, National Centre for Nuclear Research, 7 Andrzeja Sołtana Street, 05-400 Świerk-Otwock, Poland*

[2]*National Centre for Nuclear Research, A. Soltana 7, 05-400 Świerk-Otwock, Poland*

[3]*Łukasiewicz Research Network - Institute of Microelectronics and Photonics, Wólczyńska 133, 01-919 Warszawa, Poland*

*email: Magdalena.Gaweda@ncbj.gov.pl

**Abstract**

Graphite, as a material for high-temperature gas-cooled reactors (HTGR), will be exposed to harsh environment. The stability of graphite structure under irradiation is of a key importance for efficiency, reliability and security of the Generation IV nuclear reactors. Three types of nuclear grade graphite were subjected to irradiation in this research – two commercially manufactured (IG-110 and NBG-17) and the laboratory's in-home material (NCBJ). The samples were exposed to 150 keV $Ar^+$ and $He^+$ ions bombardment at 400 °C with fluences ranging from 1E12 to 2E17 ion/cm$^2$ in order to simulate in-reactor conditions. For analysis of the level of structure damage, type of created defects and crystallite size changes under ion irradiation ex-situ Raman spectroscopy was used. The methodology of spectra fitting was developed. Furthermore, SEM observation of irradiated materials was performed. Results showed structural degradation of materials by the means of amorphisation: slight at a low fluence level, rising rapidly at higher irradiation values. Furthermore, stronger structural disorder was found in the materials irradiated with heavier $Ar^+$ ions than with lighter $He^+$. Microstructural evolution of the nuclear graphites aligned with the structural deterioration in its stepwise character.

## 1. Introduction

Generation IV nuclear reactors are designed as a response to the ongoing surge in energy demand. One of them, which deserves special attention, are high-temperature gas-cooled nuclear reactors (HTGR). They are aimed to produce heat for industrial application, especially in chemical and refining industries. Their undoubted advantage is the cost, which is comparable to gas boilers. Simultaneously, they do not require fossil fuel consumption and are almost entirely $CO_2$ emission-free. Moreover, this technology ensures a high level of safety during use. In the event of any unforeseen circumstances, such as a sudden increase in temperature above a specified threshold, the chain reaction stops spontaneously, preventing the meltdown of the reactor core [1], [2].

The suitable selection of materials for HTGR is of key importance due to demanding operation conditions, including temperature up to 1000 °C, high neutron radiation (fluence up to 1E23 neutron/cm$^2$), the elongated exploitation time, high pressure (up to 100 atm) and contact with Helium as cooling gas [1], [2], [3]. Graphite, as the fundamental material used in HTGR reactors, acts as both a moderator and a construction material [4], [5], [6]. It is exceptionally attractive due to its excellent neutron moderation, high resistance to radiation, thermal and mechanical stability at high temperature. Nevertheless, the harsh operation conditions will inevitably induce significant alterations in its structure and, as a consequence, in graphite's properties. Neutron irradiation leads to changes in structure and microstructure by defects formation, dimensional changes, swelling (lattice expansion along the c-axis and shrinkage parallel to the basal plane) and, consequently, amorphisation [4], [6], [7], [8]. The type of created defects depends on the irradiation parameters (kind of ions, energy, rate, dose and temperature) and on the material characteristics (structure, crystallinity, homogeneity, and purity as the consequence of production method) [9]. Elevated temperatures in HTGRs contribute to the intricate process of defect formation and microstructural changes within graphite. The thermal conditions not only influence the kinetics of defect creation but also dictate the nature of these defects. At ambient (room) temperatures, large and stable defect structures have not been observed. As temperature rises, the mobility of atoms within the graphite lattice intensifies, affecting the rate at which defects are generated. It is commonly held, the mobility of interstitial and vacancy defects increases causing atomic rearrangement and agglomeration of point defects [10]. Previous studies reported certain types of defects typical for nuclear graphite, i.e., point defects (interstitials, vacancies), dislocations, their clusters, and dislocation loops [11], [12]. Since it has a significant impact on the operational performance of the material, the issue of structural defects evolution in different types of nuclear graphite under the reactor operation conditions requires careful analysis and consideration at the reactor design stage.

Defects formation kinetics in nuclear reactors can be simulated by irradiation with heavy ions which do not chemically change the material and can imitate neutrons, primarily with noble gases ions, i.e. Argon ($Ar^+$), Krypton ($Kr^+$), and Xenon ($Xe^+$) [4]. Furthermore, Helium ions ($He^+$) are also used to simulate damage caused by this element, since it plays the role of a cooling gas. They tend to aggregate and form round-shaped objects (the so-called helium bubbles) within the graphite material which will deteriorate mechanical properties [13].

Comprehensive analysis of structural evolution and defects characterisation can be executed using Raman spectroscopy. This method is very sensitive to structural disorder and, therefore, enables the observation of changes resulting from ion irradiation [12], [14]. In the layer structure of graphite, carbon atoms are primarily in $sp^2$ hybridisation. These bonds are observed at approximately 1580 $cm^{-1}$ in the Raman spectrum as the so-called G-band and allows recognition of regions of the graphite-like ordering. When the structure is defective, successive bands appear in the Raman spectrum, primarily, first-order bands: D and D' at 1350 $cm^{-1}$ and 1620 $cm^{-1}$, respectively. The most frequently analysed defect band is the mentioned D band due to its relatively high intensity, which significantly progresses with the amount of defects. With the growth in the intensity of the D band, the intensity of the D' band also increases. However, the D' band appears as a small shoulder at the G band. Nevertheless, with a moderate concentration of defects, the D' band can be distinguished from the G band. The level of defects and related changes in the structure can be determined based on the profile of the Raman spectrum. It is indicated by position, full width at half maximum (FWHM), intensities and ratios of the intensities of the aforementioned bands. Above all, based on the ratios of the intensity of the D, G and D' bands ($I_D/I_G$ and $I_{D'}/I_D$), it is possible to determine the degree of damage in the structure following the relationship defined by F. Tunistra and J. L. Koenig [15], [9], and the type of defects based on the dependence presented in [14]. Other identifiable group of bands are the second-order bands: D+D'', $2D_b$, $2D_a$, D+D', 2D', which occur at approx. 2460 cm$^{-1}$, 2700 cm$^{-1}$, 2720 cm$^{-1}$, 2940 cm$^{-1}$, 3240 cm$^{-1}$, respectively [14], [16]. With an increase in the number of defects, not an amplification but rather a reduction in intensity of the second order bands is observed [14]. In addition, there is a broadening of these bands. Moreover, with a certain level of defect in the structure, additional first-order bands will appear, the origin of which is not clearly defined yet. It is expected that the discrepancies will be observed between the behaviour of pristine graphites and the anticipated trends under irradiation could stem from several factors inherent to the materials themselves and the irradiation process. One potential explanation is the presence of inherent structural variations or impurities in the pristine graphite samples, which may influence their response to irradiation differently than expected. Moreover, variations in graphite composition, grain size, defect density, and crystallinity among different graphite types can lead to divergent responses to irradiation. These intrinsic material characteristics may impact the kinetics of defect formation, migration, and agglomeration, ultimately influencing the observed behaviour under irradiation.

The presented work compares changes occurring in three types of graphite: commercially available NBG-17, IG-110 and the laboratory's in-home material (NCBJ) under simulated operating conditions of nuclear reactors. The samples were subjected to Ar$^+$ and He$^+$ ion irradiation processes and examined by Raman spectroscopy. Moreover, the methodology of spectra fitting was developed. The level of induced structural damage, the type of defects formed during successive stages of irradiation and their crystallite size have been identified qualitatively. Structural examination was complemented by the microstructural analysis executed with scanning electron microscopy (SEM).

### 2. Materials and methods

In presented experiments three nuclear-grade graphites were used: IG-110 (Toyo Tanso, Japan), NBG-17 (SGL Carbon, Germany), NCBJ (unused replacement material, intended for the decommissioned EWA reactor at the National Centre for Nuclear Research, Poland). IG-110 graphite was subjected to irradiation and further examination in the dimensions provided by the manufacturer. NBG-17 and NCBJ graphites required preparation. They were cut into slices approximately 5 mm thick. Additionally, to address concerns regarding the influence of surface condition on Raman spectra, a meticulous sample processing methodology was employed. This methodology involved the optimization of surface preparation processes to mitigate any potential interference from polishing materials within the porous structure of the graphite. Initially, the polishing process utilized sandpaper with a 4000-grit size (corresponding to SiC particles of approximately 5μm) followed by rinsing in ethanol using an ultrasonic cleaner for 3 to 5 minutes. Subsequently, polishing with polishing paper continued until surface contaminants were effectively eliminated. Further preparation of the surface were achieved through the application of a broad beam of low-energy Ar+ ions at 4 keV for a duration of 50 minutes. This ion beam polishing process was executed using the Hitachi IM4000 device dedicated to ion beam polishing of samples and creating cross-sections using low-energy Ar+ ions. This step facilitated the attainment of a high-quality surface while eliminating residual polishing material and damage remnants from the initial surface preparation stage. These rigorous preparation techniques were instrumental in ensuring consistency and minimizing variations in Raman spectra across different locations on the graphite samples.

Materials were irradiated with ions to simulate the formation of radiation defects in the reactor environment. For this purpose, Ar+ ions were used to simulate the neutrons to which graphite is exposed during operational conditions in the reactor, while He+ ions were utilised because He+ serves as a gaseous coolant in HTGR reactors. The irradiations were performed at 400 °C and the energy of ions of 150 keV. The parameters of irradiation and the level of damage in the material (displacements per atom, dpa) were presented in Table 1. In order to estimate the value of dpa level, the simulations were carried out using the SRIM (Stopping and Range of Ions in Matter) software to investigate the influence of the irradiation on the graphites [17]. The significance of SRIM input parameters lies in providing ability to quantify the extent of structural changes induced by irradiation in different graphites. Fluence provides information about the intensity of ion bombardment experienced by the sample, while dpa offers a measure of the density of atomic displacements within the material. The aim of applying various fluence values, gradually increasing, was to simulate the behaviour of graphites in the reactor over increasingly longer periods of operation. By correlating these values with the evolution of Raman spectra profiles, we can elucidate the relationship between irradiation parameters and structural modifications in graphitic materials. The irradiation energy value was dictated by ensuring the appropriate depth of the dpa peak.

*Tab. 1. Ar+ and He+ ion irradiation fluence with dpa (displacements per atom) estimated with SRIM software*

| Fluence [ion/cm$^2$] | Estimated displacements per atom (dpa) | | Estimated depth of dpa peak [µm] | |
|---|---|---|---|---|
| | Ar+ | He+ | Ar+ | He+ |
| 0 | - | - | - | - |
| 1E12 | 0.00129 | 0.00005 | 0.08 | 0.63 |
| 3E12 | 0.00387 | 0.00015 | | |
| 1E13 | 0.01292 | 0.00051 | | |
| 3E13 | 0.03875 | 0.00152 | | |
| 1E14 | 0.01292 | 0.00505 | | |
| 3E14 | 0.38475 | 0.01516 | | |
| 1E15 | 1.29162 | 0.05052 | | |

| 3E15 | 3.87487 | 0.15156 | | |
| 1E16 | 12.91624 | 0.50519 | | |
| 3E16 | 38.74872 | 1.51556 | | |
| 1E17 | 129.16239 | 5.05188 | | |
| 2E17 | 258.32477 | 10.10376 | | |

Changes in the studied materials' structure resulting from irradiation were observed ex-situ using a confocal Raman microspectrometer WITec alpha 300R (Oxford Instruments, UK), equipped with a CCD camera. Samples were excited with 532 nm Nd:YAG laser with 2.4 mW power applied to avoid inducing additional structural changes and surface damage. Spectra were collected using x100 lens in the range from 100 to 3600 cm$^{-1}$ with 30 s acquisition time operated by the Control 5.2 software. Multiple point measurements were conducted on the samples, facilitating the selection of the most representative spectrum reflecting the impact of ion implantation on the sample with ease. The first stage of data processing covered the cosmic ray removal (CRR) and the baseline correction, using polynomial functions in the WITec Project FIVE 5.2 PLUS software. Then the spectra were exported and further analysed in Python environment. In order to accurately determine the parameters of the spectrum, Voigt functions were fitted to individual bands, the convolution of which coincided with the shape of the obtained Raman spectra [18]. The adopted fitting assumptions resulted in 97% compliance level, calculated using the least square method. To ensure an adequate and repeatable fit, the range of wavenumbers corresponding to the occurrence of individual bands was determined by referencing literature values and expanded to accommodate potential shifts of these bands. Accordingly, the D band must fall within the range of 1300 1420 cm$^{-1}$, the G band within 1500-1610 cm$^{-1}$ and the D' band within 1600 1650 cm$^{-1}$[19]. Exemplary spectra with fitted functions are presented in Figure 1. For non-irradiated samples and low doses (Fig. 1a), three Voigt functions were fitted to first-order bands (range between 1300 to 1700 cm$^{-1}$) and five Voigt functions to second-order bands (2400 to 3300 cm$^{-1}$) [20], [21]. However, the analysis of spectra of highly defective samples (e.g., irradiated at fluences higher than 3E13 ion/cm$^2$, Fig. 1b-c), demanded taking into the account two additional bands, D* and D**, in the fitting process. Therefore, within the range of first-order Raman bands (between 1300 and 1700 cm$^{-1}$), five Voigt functions were applied. According to findings in the literature, the D** band originates from amorphous graphite and typically appears between the D and the G bands, thus the range was limited to 1450-1520 cm$^{-1}$. On the other hand, the D* band can be attributed to graphite lattice defects, sp$^3$ vibrations, or the presence of oxygen groups, and typically appears in range of 1000 -1300 cm$^{-1}$[21], [22], [23], [24]. For the most defective samples (such as those irradiated at 2E17 ion/cm$^2$), it was impossible to achieve a good fit for the second-order bands due to their enormous width, very low intensity and overlapping.

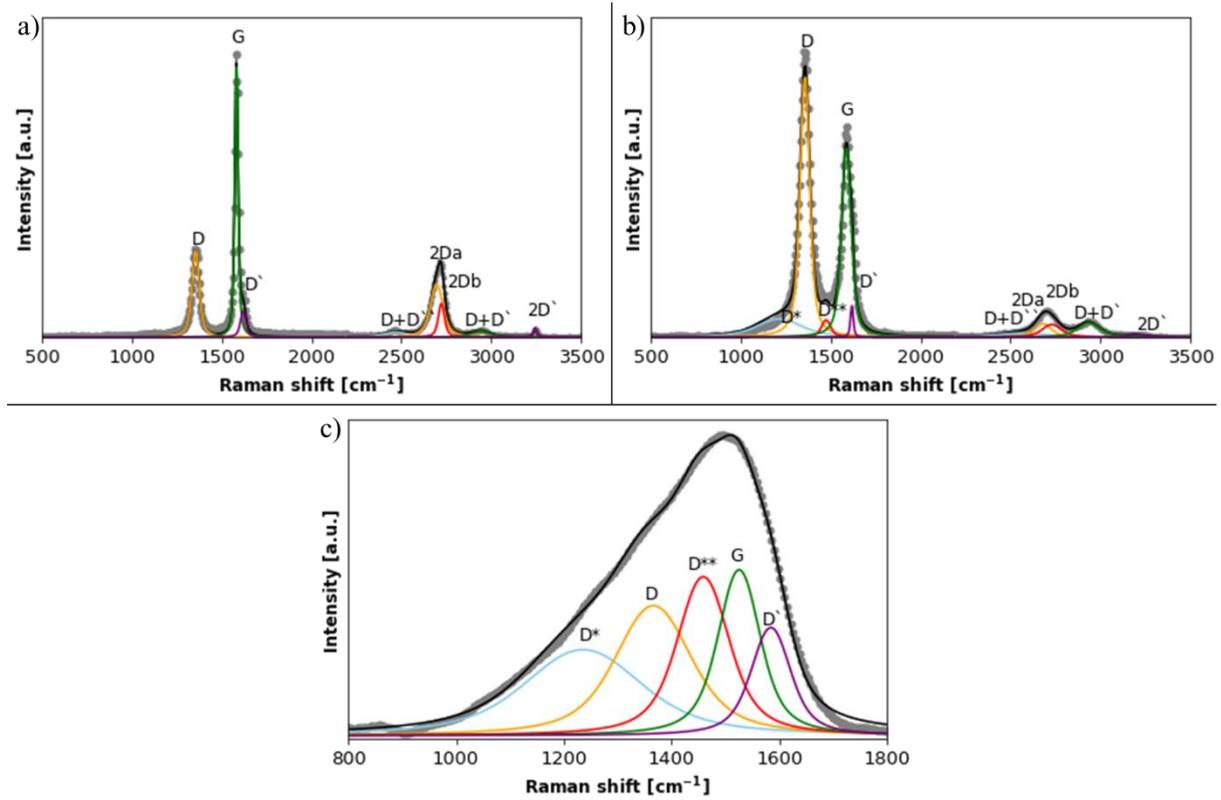

*Figure 1. Exemplary Raman spectra of IG-110 graphite irradiated with Ar+ ions with fluence of 1E12 ion/cm² (a), 3E13 ion/cm² (b) and 2E17 ion/cm² (c) fitted with Voigt function, presented with bands assignment.*

The bands' parameters, obtained from the analysis of spectra in Python environment, were used for presentation of factors indicating structural disorder in the studied materials [9], [20]. Plots presented in Figures 2-7 were created in OriginPro 9.9. Charts include positions, FWHM, $\frac{I_{D'}}{I_G}$, $\frac{I_D}{I_G}$ and $\frac{I_{D'}}{I_D}$ ratio, where $I_D$, $I_G$ and $I_{D'}$ represents the maximum intensities of the D, G and D' bands, respectively.

Moreover, the size of crystallites was analysed. Calculations for pristine and irradiated samples were performed according to Eq. 1 [20], where values for phonon dispersion and decay length parameters values were assumed to be 195 and 32 nm, respectively [20].

$$L_a = \frac{l_C}{2} \ln\ln\left[\frac{C}{\Gamma_G^A(L_a) - \Gamma_G^A(\infty)}\right] \qquad \text{(Eq. 1)}$$

where:

$C$          is a parameter related to the phonon dispersion $\omega(q)$,

$l_C$          is the full decay length [nm],

$\Gamma_G^A(L_a)$     is FWHM (full width at half maximum) of the G band of the given spectra,

$\Gamma_G^A(\infty)$     is the minimum FWHM of the G band for a pristine sample.

SEM observations were conducted to morphologically characterise NBG-17, IG-110, and NCBJ graphite samples using ThermoFisher Scientific™ Helios™ 5 UX Scanning Electron Microscope. The specimens were subjected to microstructural observations before and after irradiation with Ar+ and He+ ions. Employing an in-chamber electron (ICE) detector, imaging encompassed various magnifications, enabling the discernment of morphological distinctions within the materials. The electron energy used during the analysis was 5 keV, with a 0.8 nA probe current.

## 3. Results and discussion

Raman spectra of nuclear graphites before and after irradiation were presented in the form of topographic plots (Fig. 2). The OY axis represents rising radiation dose, whereas, the colour evolution indicates intensity. Before plotting, spectra were normalised for easier evaluation. Structural evolution of the material is clearly visible by increase of the intensity of the D band approx. at 1350 cm$^{-1}$, shift of both D and G bands, their broadening up to the point where only one broad complex band is visible. This is assisted with second-order bands fading. Moreover, these effects may be observed at lower value of fluence for the heavier Ar+ ions (Fig. 2a), whereas structural deterioration for lighter He+ ions irradiation requires a significantly higher dose (Fig. 2b). The fluence and dpa corresponding to the key points of Raman spectra profile evolution are gathered in Table 2. It may be noticed, that the significant broadening of the D and G bands and further combining of the bands (associated with amorphisation) occurs at significantly higher dose and dpa for He+ than for Ar+, with the exception of IG-110 graphite.

*Tab. 2. Values of fluence [ion/cm$^2$] and displacements per atom (dpa) [-] corresponding to the crucial Raman spectra profile changes.*

|  | **NBG-17** | | **IG-110** | | **NCBJ** | |
| --- | --- | --- | --- | --- | --- | --- |
|  | fluence | dpa | fluence | dpa | fluence | dpa |
| **Ar+** | | | | | | |
| significant increase of intensity and broadening of D band | 1E14 | 0.01292 | 1E14 | 0.01292 | 1E14 | 0.01292 |
| D and G bands combining | 1E15 | 1.29162 | 3E15 | 3.87487 | 3E15 | 3.87487 |
| **He+** | | | | | | |
| significant increase of intensity and broadening of D band | 3E16 | 1.51556 | 3E16 | 1.51556 | 3E16 | 1.51556 |
| D and G bands combining | 1E17 | 5.05188 | 3E16 | 1.51556 | 1E17 | 5.05188 |

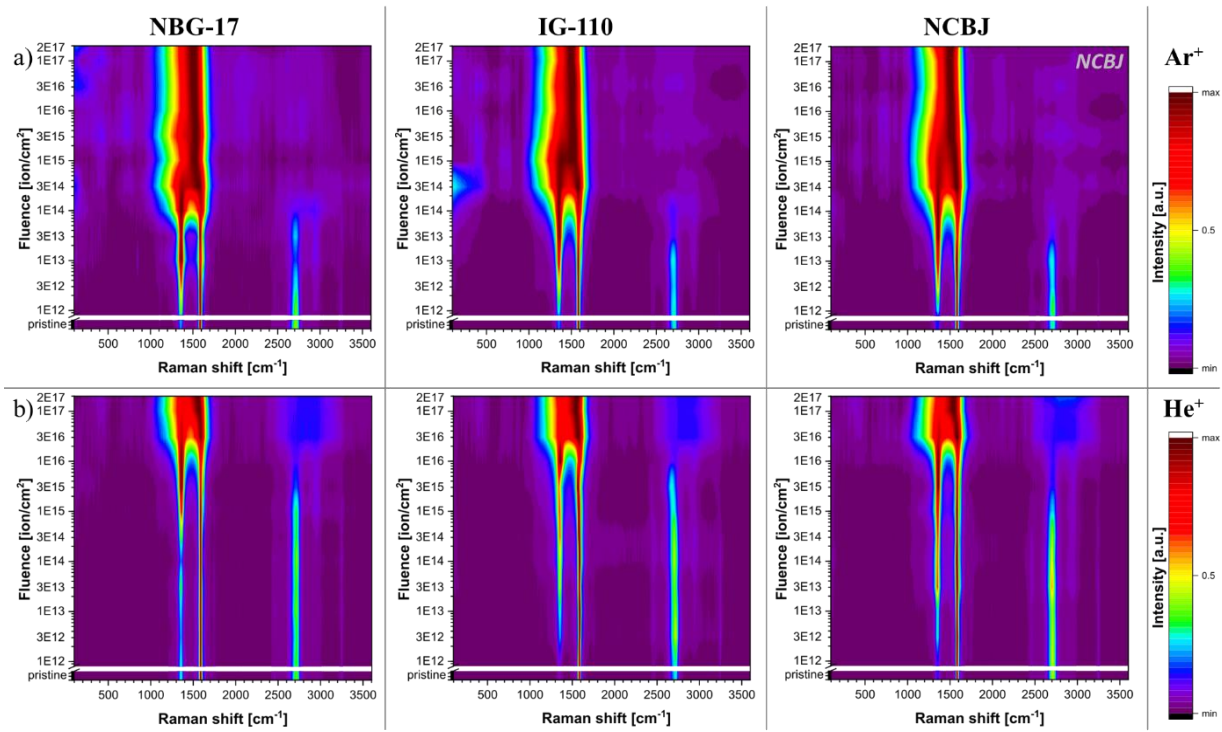

*Figure 2. Topographic plots of Raman spectra evolution of nuclear graphites under irradiation with Ar+ (a) and He+ (b) ions.*

The results of the positions, intensity and FWHM of Raman bands (particularly G, D and D') were obtained by fitting the Voigt function to the individual spectra in the Python environment for three types of graphite at every fluence (Fig. 3-4). By comparing them, one can observe that the tendency of spectral changes with increasing fluence has the same character. For low doses, the D and G bands can be clearly distinguished. The graphite-derived G band is narrower and has higher intensity than the D band. However, with an elevation in fluence, the relative intensity of the D band experiences a noticeable surge. With higher fluence, analysed bands become broader. Moreover, a shift of the position of the D band towards higher values of wavenumbers can be clearly observed. The increase in intensity of the D band at approximately 1350 cm$^{-1}$, along with the shift and broadening of both the D and G bands in the Raman spectrum of irradiated graphite, suggests significant structural alterations in the material. Specifically, the D band intensity enhancement often indicates the presence of defects, disorder, or the introduction of sp$^3$-hybridized carbon domains within the graphite lattice due to ion irradiation. These alterations indicate an increasing level of defects and disorder between the graphite layers after irradiation, accompanied by a diminution of crystallite size towards nanocrystalline graphite. Furthermore, the shift and broadening of the D and G bands imply changes in the carbon bonding environment and the formation of disordered carbon structures, possibly arising from lattice distortions, or vacancies, which contribute to the further amorphisation of the structure induced by irradiation. Nevertheless, position fluctuation of the G band is more complex: it remains stable or slightly rises with low irradiation dose, and after reaching a threshold value shifts strongly towards lower wavenumbers. The dynamics of these effects differ among the used ions. In case of Ar+ ion irradiation, broadening and, hence, overlapping of the D and G bands occurs above the fluence of 3E14 ion/cm². Changes in Raman spectra obtained from graphites irradiated with He⁺ ions begin to occur at higher fluence of 1E16 ion/cm². Up to this level, the D' band and the G band are clearly distinguishable, and second-order bands can also be decomposed from the spectra profile. As the positions of the D and G bands shift, and their FWHM widen, they consequently overlap. It indicates an increasing level of defects and disorder between the graphite layers after irradiation, diminish of size of crystallites towards nanocrystalline graphite (detailed description in the following sections of the article) and further amorphisation of the structure [25], [26]. The observed discrepancies in spectra of samples implanted with different ions result from size and mass differences of used ions. Irradiation with He⁺ ions, which are much lighter than Ar⁺ ions, introduces a smaller amount of defects. This is also confirmed by the 25 times higher number of displacements per atom (dpa) after irradiation with Ar⁺ compared to He⁺ (Table 1). Despite the diverse paths and kinetics of defect formation, at the highest fluence, the level of structural damage is similar for both types of bombarding ions. This similarity is evidenced by comparable values in both the positions (Fig. 3) and FWHM (Fig. 4) of the G band.

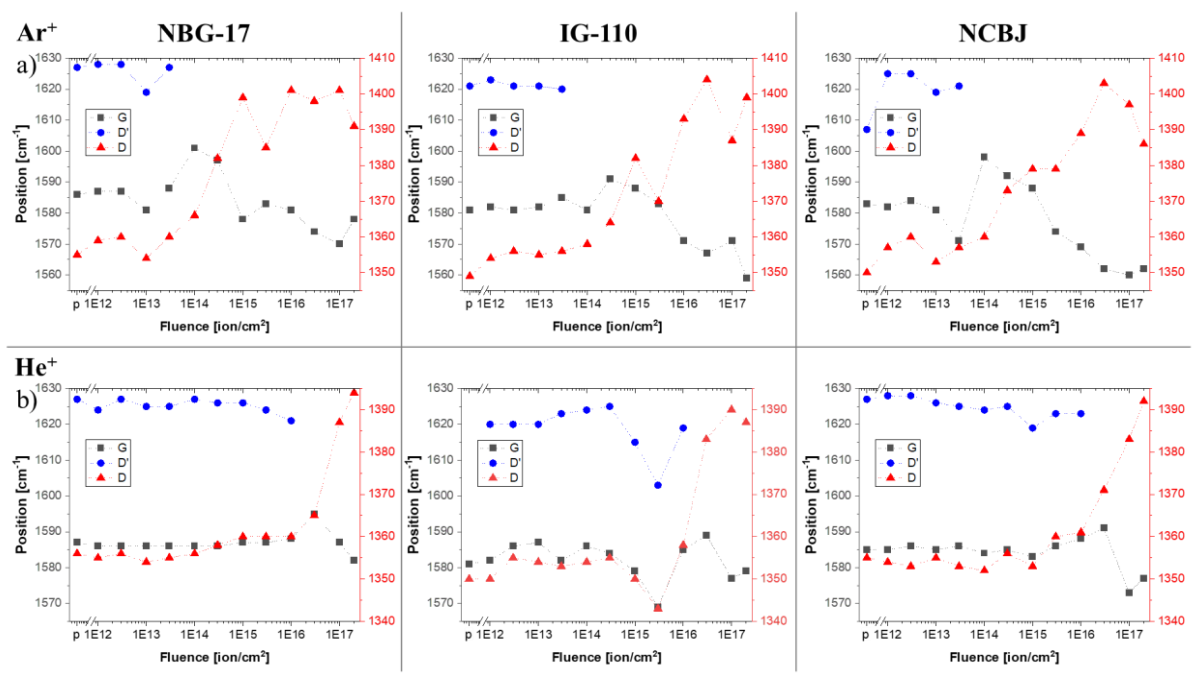

*Figure 3. Plot of G, D and D' bands positions as a function of irradiation fluence for samples irradiated with Ar⁺ (a) and He⁺ (b) ions.*

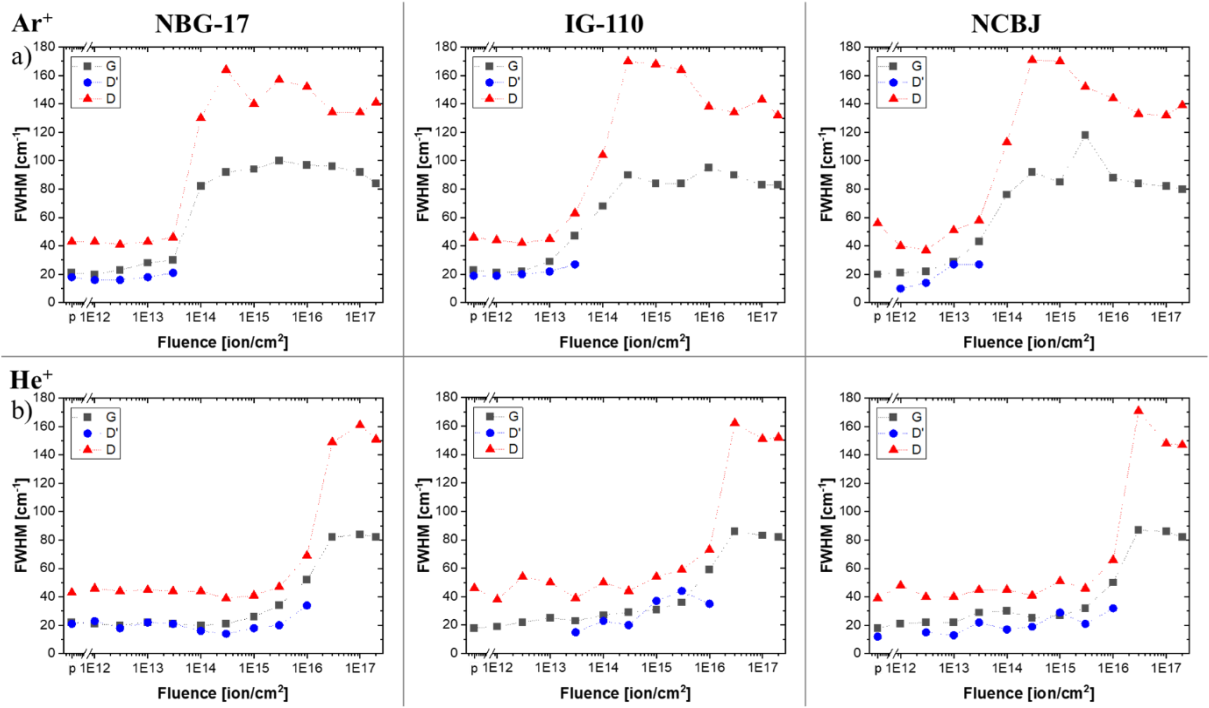

*Figure 4. Plot of the full width at half maximum (FWHM) of G, D and D' bands as a function of irradiation fluence for samples irradiated with Ar⁺ (a) and He⁺ (b) ions.*

The intensity ratio of the D to D' bands was employed to determine the types of defects present in material. According to the literature [9], as the number of defects in the material builds up the intensity of the D and D' bands rises as well. The ratio of $\frac{I_D}{I_{D'}}$ can be used to obtain information about the nature of defects. Specifically, it was determined that this ratio is approximately 7 for vacancy-like defects, 3.5 for boundary-like defects and 13 for defects associated with sp³ hybridisation [9]. Intermediate values indicate the presence of mixed type of defects. Based on the ratio of $\frac{I_D}{I_{D'}}$ for particular samples, the type of defects and their evolution with ion irradiation fluence were determined (Fig. 5). It should be emphasised that the presented graphs were generated only for fluences at which the determination of the D' value was possible (for samples with fluence smaller than 1E14 ion/cm² for Ar⁺ and 3E16 ion/cm² for He⁺).

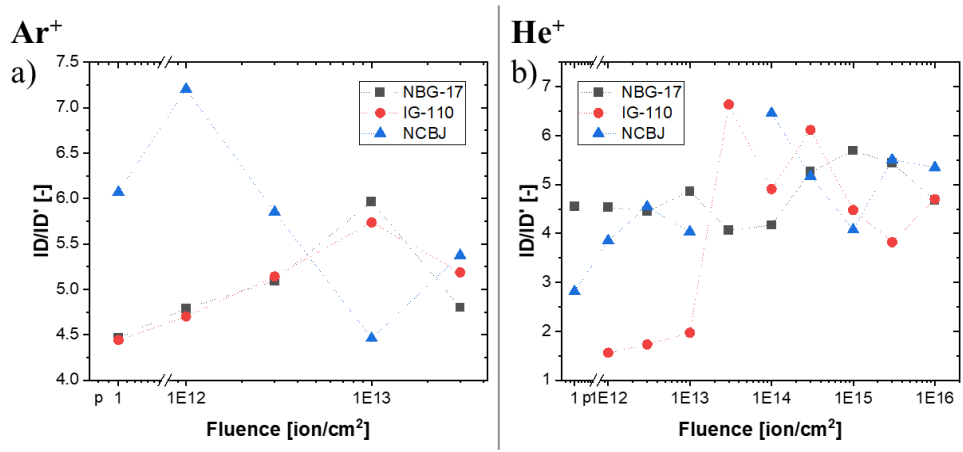

*Figure 5. Dependence of the $\frac{I_D}{I_{D'}}$ ratio on the irradiation dose for Ar⁺ (a) and He⁺ (b) ions.*

The evolution of the $\frac{I_D}{I_{D'}}$ ratio becomes apparent already at low fluences of Ar⁺ ion (Fig. 5a) and it can be stated that irradiation with Ar⁺ ion introduces both types of defects: vacancies and edge defects. The presented results suggest differences in the types of defects in non-irradiated graphite samples. For pristine IG-110 and NBG-17, boundary-like defects ($\frac{I_D}{I_{D'}} \approx 4.5$) were prevalent. The presence of different types of defects in non-irradiated graphite samples, such as boundary-like defects in IG-110 and NBG-17, and vacancy-like defects in NCBJ graphite, has significant implications for the overall structural stability of the materials. Boundary-like defects, typically observed in IG-110 and NBG-17, may indicate the presence of structural imperfections at the boundaries of crystalline regions. These defects could potentially weaken the material's structural integrity, making it more susceptible to damage under irradiation. On the other hand, vacancy-like defects prevalent in NCBJ graphite suggest the presence of vacancies within the graphite lattice, which can lead to localized distortions and dislocations. While these defects may not inherently compromise the material's structural stability, their presence could facilitate the nucleation and propagation of damage under irradiation. As the fluence increased, the point defects started to be created and become predominant in these materials at fluence 1E13 ion/cm². However, at higher fluence this ratio started to decrease, suggesting that vacancies began to agglomerate, forming dislocation loops and increasing the amount of edge defects [27]. In contrary, for NCBJ graphite, the vacancy-like defects were prevalent ($\frac{I_D}{I_{D'}} \approx 6$) for a non-irradiated sample. Similarly like in case of the other materials, initially the number of vacancies multiplies. However, the evolution of the vacancies into clusters and dislocation loops took place at lower doses. This suggests that the process of defects generation and their evolution is much more dynamic in NCBJ graphite than in other studied materials. This indicates the less stable and more defective structure of the NCBJ graphite compared to other materials.

In the case of He⁺ irradiated graphites (Fig. 5b), the evolution of the $\frac{I_D}{I_{D'}}$ ratio with increasing fluence is more gradual. This is associated with the slower damage rate of the material at increasing fluences, attributing to the smaller mass and size of Helium atoms. Moreover, the dispersion of value in this ratio is much greater, indicating the continuous evolution of defects. The analysed materials exhibit a similar trend of changes as observed under Ar⁺ irradiation: vacancies appears initially ($\frac{I_D}{I_{D'}}$ ratio raises up to 7) and then they start to agglomerate, forming dislocation loops ($\frac{I_D}{I_{D'}}$ ratio drops). Then this process repeats. However, pristine graphites show discrepancies from the previous analysis, suggesting the inhomogeneity of the measured materials. Nevertheless, this issue does not affect the observed trend of changes. Furthermore, the creation of vacancies takes place at higher fluences compared to Ar⁺ ions and varies among the different materials. NBG-17 reaches the maximum number of vacancies later than the other samples, and the evolution of the $\frac{I_D}{I_{D'}}$ ratio is smoother, indicating the most stable structure among the studied materials.

A broader analysis of the level of damage in the structure in each of the nuclear graphites was conducted based on the relationship between the intensity of $\frac{I_D}{I_G}$ and the applied fluence of irradiation (Fig. 6). According to the work [28], in the case of a small defect concentration, when the average distance between the atoms is large, ion bombardment reduces the dimensions of the ordered structure, but does not significantly affect the number of hexagonal rings. This causes the intensity of the D band to rise with increasing fluence. On the other hand, at high fluence, if the long-range order is reduced to a short-range order with dimensions comparable to those of hexagonal rings, a significant reduction in the D band intensity is expected. These results confirm that for $Ar^+$ ion irradiation a lower number of defects was formed in NBG-17 at low fluences (to the value of 1E14 ion/$cm^2$) than in the case of NCBJ and IG-110 graphite. For higher doses, the amount of defects in all three types of graphite is similar. In case of $He^+$ ions, the $\frac{I_D}{I_G}$ ratio in NBG-17 graphite begins to enlarge at higher fluences, so defects occur at higher fluences compared to other graphite samples. However, at high fluences, the $\frac{I_D}{I_G}$ ratio converges to a similar value for all types of graphite. This indicates a lower impact of low fluence of $He^+$ ions on the amount of structural defects in the case of NBG-17 graphite than in other samples, and a similar degree of defect for high fluences in all types of samples. Referring the obtained information to the results after irradiation with $Ar^+$ ions, it can be seen that in all samples the level of defects escalates faster (at lower fluences) and finally reaches a higher value than for $He^+$ irradiation. This is in line with the simulated structural failure (dpa), which is significantly lower for $He^+$ than $Ar^+$ ion irradiation. $He^+$ ions, being lighter, introduce a smaller amount of defects per ion impact compared to $Ar^+$ ions, resulting in a delayed onset of structural changes. This is attributed to the greater energy deposition per ion collision, leading to the formation of larger defects and increased structural disorder within the graphite lattice [29]. Based on the graph, it can be concluded that after $He^+$ irradiation loss of crystallinity is achieved by NBG-17 graphite at the latest.

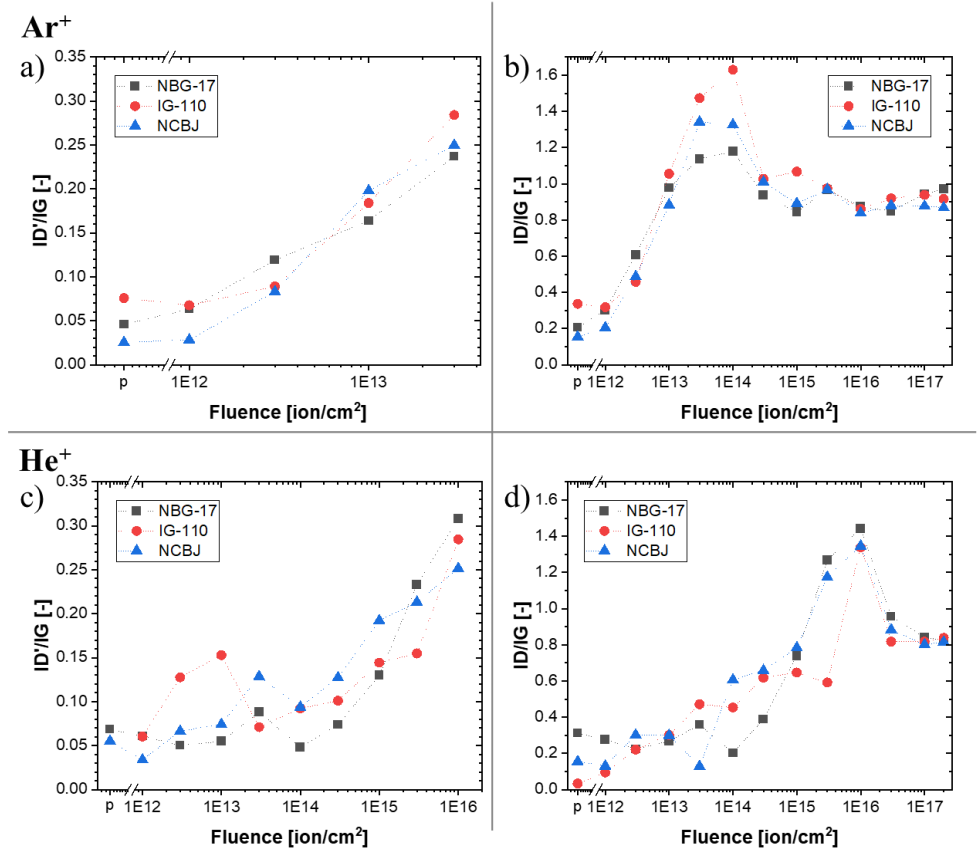

*Figure 6. Plot of the $\frac{I_{D'}}{I_G}$ (a, c) and $\frac{I_D}{I_G}$ (b, d) intensity ratios as a function of irradiation fluence after irradiation with Ar⁺ (a, b) and He⁺ (c, d) ions.*

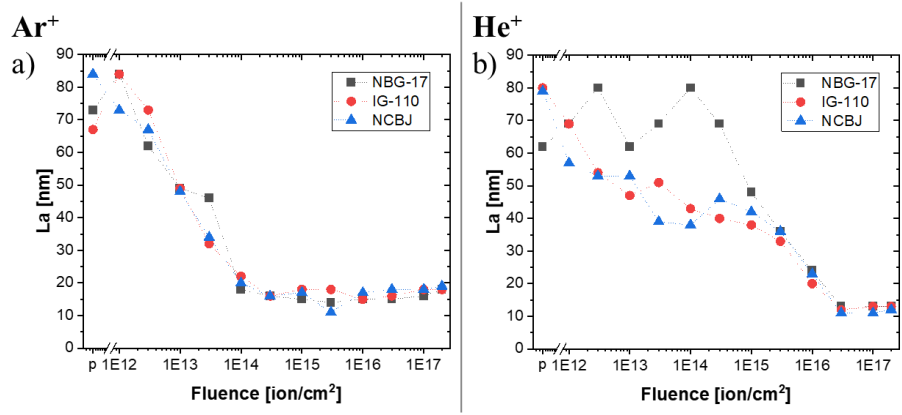

*Figure 7. Evolution of size of crystallites after irradiation with Ar⁺ (a) and He⁺ (b) ions.*

The evolution of nuclear graphite structural disorder after ion irradiation can be assessed by depicting the relationship between the crystallite size and fluence (Fig. 7). Analysing the results obtained for Ar⁺ ions irradiation (Fig. 7a), the minimal size of crystallites is achieved at a fluence of 1E14 ion/cm² and higher. Prior to this point, a steady linear decrease can be observed from the lowest fluence. This implies a rapid transition of the material into nanocrystalline form and subsequent amorphisation. This indicates that the damage induced by irradiation leads to the fragmentation and dissolution of crystalline regions within the graphite lattice, resulting in the formation of smaller crystallites and eventual loss of long-range order. In contrast, He⁺ (Fig. 7b) irradiation resulted in a stepwise plot of the La curve, indicating a delay in the materials' transition into nanocrystalline graphite and amorphisation occurring at a fluence of 3E16 ion/cm². This indicates path and rate of amorphisation, consistent with other presented results. However, NBG-17 exhibits the largest crystallite size up to fluence of 1E15 ion/cm². This also confirms previous results, that NBG-17 presents the slowest loss of crystallinity. Despite differences in the dynamics, the final size of crystallites is coherent for all three types of graphite under irradiation with both ions. Although the calculation of crystallite size in Figure 7 may still yield a value at the highest damage level, it is important to note that this measure represents the size of the remaining crystalline domains rather than the overall structural state of the material. Even in heavily damaged samples, there may still be localized regions or domains with some degree of crystallinity, which can contribute to the observed crystallite size measurement. However, the broadening and merging of the D and G bands indicate that the majority of the material has lost its long-range order and has transitioned into an amorphous or nanocrystalline state, despite the presence of residual crystalline domains.

SEM observations covered pristine samples and materials after irradiation (Fig. 8-10). The focus was put on the fluence values where crucial structural and morphology changes were induced by ions. Figure 8 presents SEM images of pristine nuclear graphites, namely NBG-17, IG-110, and NCBJ, at low and high magnification. Clear differences in morphology are evident among these materials. At lower magnifications, NBG-17 exhibits the lowest surface roughness, followed by IG-110 and NCBJ with increasing roughness. Upon closer examination at higher magnifications, the pristine NBG-17 sample possesses a distinct microstructure consisting of numerous small, compact flakes. IG-100 presents a microstructure characterised by large, flat areas, while NCBJ graphite exhibits more pronounced surface heterogeneity.

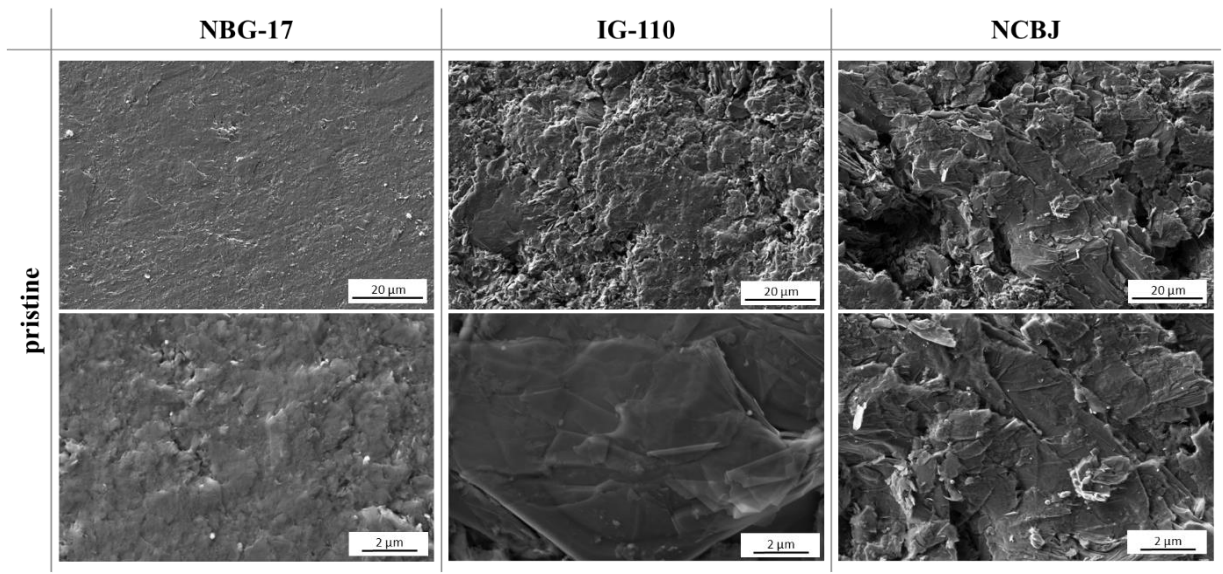

*Figure 8. SEM images of the pristine nuclear graphites.*

In Figure 9, where nuclear graphites after irradiation by Ar⁺ ions are shown, graphite samples reveal microstructural heterogeneity with increasing value of irradiation fluence. At a fluence of 3E14 ion/cm², NBG-17 shows the emergence of round-shaped objects, a phenomenon less pronounced in other samples. The round-shaped objects observed in the NBG-17 graphite samples at a fluence of 3E14 ion/cm² are likely pores resulting from ion irradiation. When energetic ions penetrate the graphite lattice, they can induce the creation of defects, including vacancies and interstitials. The accumulation and coalescence of these defects can lead to the formation of voids or pores within the material. At a higher fluence of 1E17 ion/cm$^2$ in NBG-17 graphite, these objects begin to increase slightly but maintain a compact and round shape. Notably, IG-110 and NCBJ display larger corrugations than NBG-17, at a lower fluence value, a discernible degradation in microstructural integrity becomes evident. Upon morphological examination of the IG-100 sample at reduced magnification, regions devoid of round-shaped objects persist, whereas the NCBJ sample exhibits comprehensive coverage with corrugations formation. At the pinnacle implantation dose of 2E17 ion/cm$^2$, discernible characteristics include the compact and more spherical nature of round-shaped objects within the NBG-17 sample, distinguishing it from other graphite samples. While NCBJ graphite manifests the most substantial alterations in microstructure, NBG-17 graphite emerges as the most resilient, corroborating findings derived from Raman spectroscopy analyses.

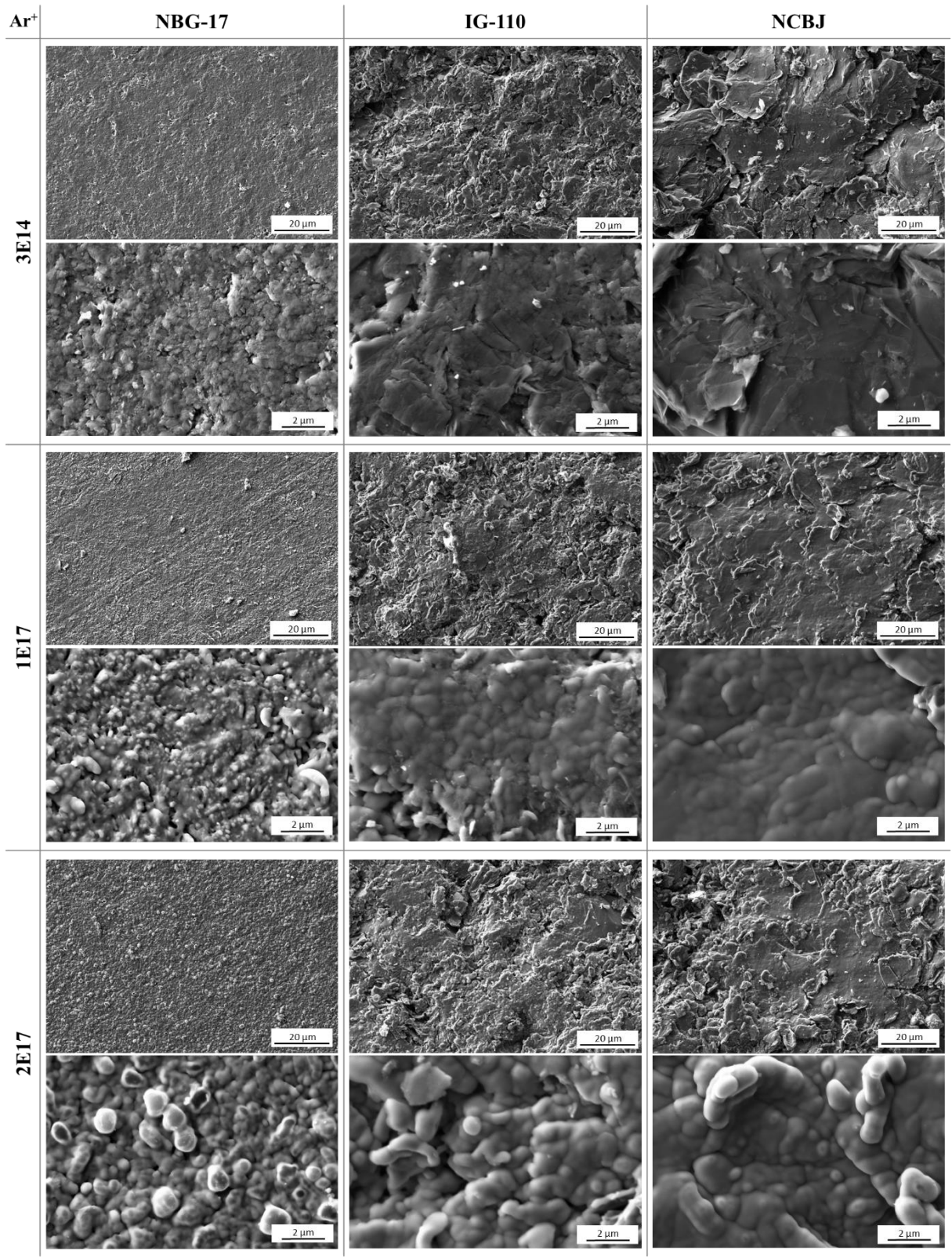

*Figure 9. SEM images of nuclear graphites under irradiation by Ar+ ions with fluence of 3E14 ion/cm², 1E17 ion/cm² and 2E17 ion/cm² corresponding to the crucial morphology changes.*

Figure 10 illustrates key features of morphology changes of nuclear graphites irradiated with various irradiation conditions. He+ irradiation at 3E16 ion/cm² does not induce significant alterations in the microstructure of NBG-17 graphite compared to the pristine sample. In IG-110, small pores become visible, while both, small pores and changes in edges appearance towards a more rounded form for NCBJ graphite. At the highest fluence of 2E17 ion/cm$^2$, NBG-17 maintains the most stable microstructure, with smaller round-shaped objects comparing to other graphite samples, bearing some similarities to the pre-implantation morphology. In contrast, IG-110 and NCBJ samples exhibit microstructures entirely covered with round-shaped objects of similar size and morphology.

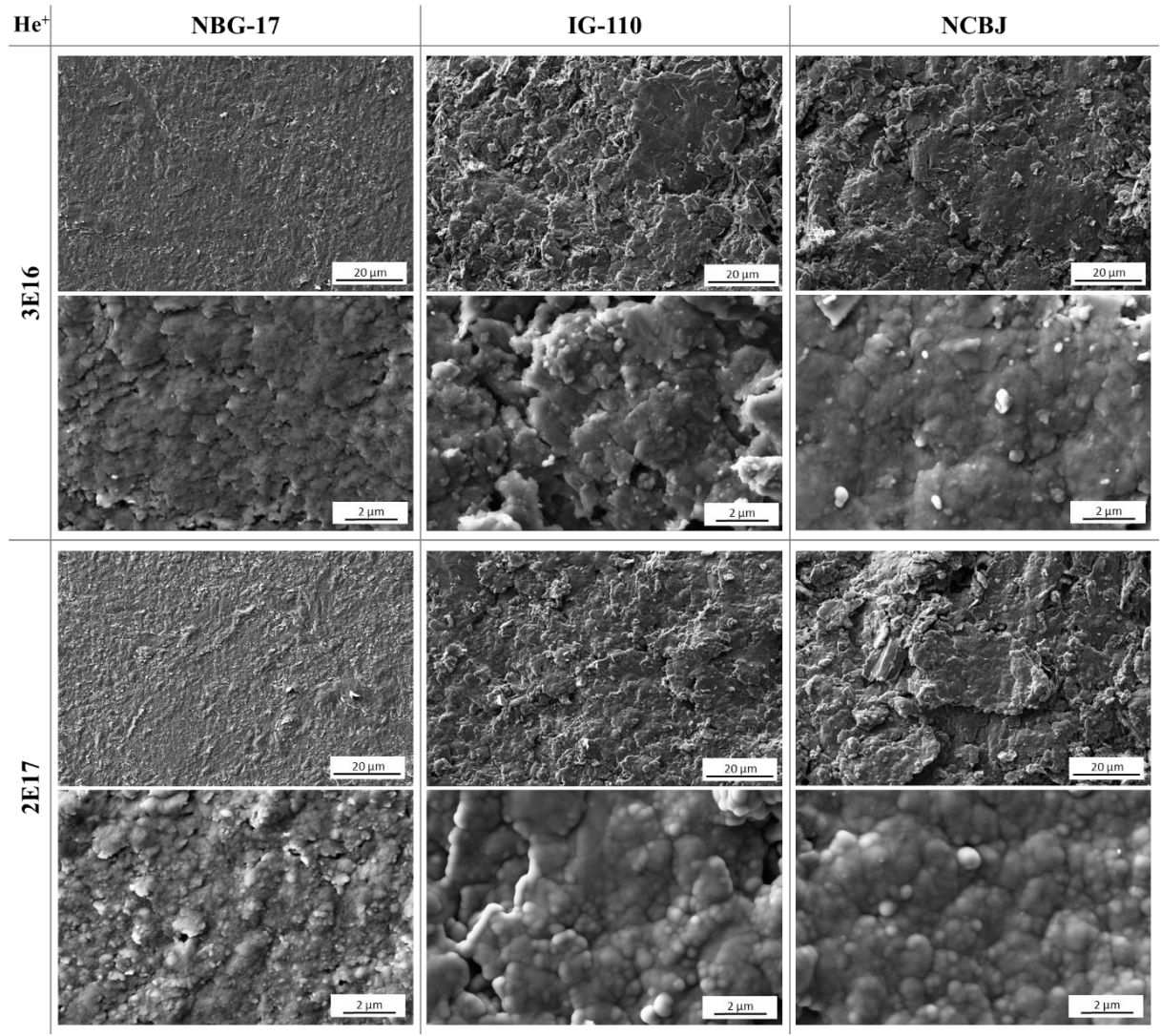

*Figure 10. SEM images of nuclear graphites under irradiation by He$^+$ ions with fluence of 3E16 ion/cm$^2$ and 2E17 ion/cm$^2$ corresponding to the crucial morphology changes.*

SEM observations complement the results obtained from Raman spectroscopy by providing visual confirmation of the structural changes identified spectroscopically. In assessing the impact of Ar$^+$ and He$^+$ ions on material properties, notable distinctions emerge. Specifically, the heavier Ar+ ions induce discernible structural alterations at lower ion fluences compared to He+ ions. This discrepancy stems from the greater mass and energy deposition of Ar+ ions, resulting in more pronounced damage to the sample. Lower energy deposition of He+ ions leads to localized damage and smaller cluster formation, contrasting with the more gradual defect accumulation observed in IG-110 and NBG-17 displacement cascades within the lattice. This results in fewer defects and less pronounced disorder compared to Ar+ ions. Moreover, SEM observations of graphite samples irradiated with He+ ions often show milder morphological alterations, such as the formation of small pores and subtle changes in surface morphology, compared to samples irradiated with Ar+ ions. Observed structural changes, such as the emergence of larger round-shaped objects in case of Ar$^+$ ions, align with findings from Raman spectroscopy. The emergence of the objects observed in SEM images corresponds to the increased intensity of the D band in Raman spectra, indicating the presence of defects density and disorder induced by ion irradiation. Moreover, the variation in surface roughness observed in SEM images aligns with the broadening and shifting of Raman bands, reflecting changes in the graphite's carbon bonding environment and the formation of disordered carbon structures.

## 4. Conclusions

This work compares the structural changes in three types of nuclear graphite: NBG-17, IG-110 and NCBJ after irradiation with Ar$^+$ and He$^+$ ions, using the same irradiation conditions. Measurements were made utilising a confocal Raman spectrometer. The Python script was developed to accurately fit Voigt functions to the Raman bands with a compliance level of $R^2 \sim 0.9970$. Based on this, the exact quantitative parameters of the bands, such as G, D and D', were extracted, and a detailed analysis of the structural changes resulting from irradiation with two types of ions was performed. First of all, the level of structural disorder and the types of defects formed during the process were determined. Additionally, crystallite size was calculated for different fluences. The comprehensive analysis enabled a comparison of the behaviour of various types of graphite in a simulated nuclear reactor environment.

It was determined that Ar$^+$ irradiation induces a higher level of damage in the structure compared to He$^+$, as implied by 25 times higher dpa (calculated through simulations in the SRIM program), and a higher D band to G band intensity ratio. Moreover, under irradiation with He$^+$ ions, the evolution of structure began at higher fluences compared to irradiation with Ar$^+$. In addition, the dependence of the increasing FWHM with the increasing ion fluence, was observed for both types of ions used for irradiation. This was an evidence of progressive amorphisation. The presented analysis revealed few differences among the examined materials, clearly visible for lower fluences. The materials chosen for experimentation were initially assumed to possess similar structural characteristics prior to ion implantation. However, notable variations in their response to implanted ions were observed, indicating potential differences in their intrinsic properties. These differences, likely stemming from slight discrepancies in the manufacturing process of pristine graphite samples, may exert varying influences on their susceptibility to irradiation-induced alterations, deviating from anticipated trends. Differences in the types and rates of defect formation manifest across the graphite types under ion irradiation. Variations in defect types, such as boundary-like defects in IG-110 and NBG-17 versus vacancy-like defects in NCBJ graphite, suggest diverse mechanisms of defect generation and evolution. The dynamic response of NCBJ graphite, characterized by rapid defect evolution and cluster formation, contrasts with the more gradual defect accumulation observed in IG-110 and NBG-17. The defects generation was most dynamic for NCBJ graphite (as observed under Ar$^+$ ions). Furthermore, the analysis of evolution of defects type and level with fluence indicates that the structure of NBG-17 is the most stable, showing the smallest changes in defects evolution and the slowest loss of crystallinity as evidenced by larger crystallite sizes at comparable fluences, IG-110 and NCBJ graphite display more rapid decreases in crystallite size, indicative of pronounced structural alterations. Moreover, for this type of graphite, defects appear at higher fluences. On the basis of these findings, it was determined that NBG-17 graphite remains most stable under low fluence ion irradiation and has been found as the most resistant to irradiation among the studied materials. At high fluences, no significant differences were observed between used graphites. The aforementioned findings were complemented through morphology examination employing a scanning electron microscopy. Distinct morphological differences were identified among the pristine graphite samples, with NBG-17 displaying flake-like microstructure of the lowest surface roughness among samples. Implantation with $Ar^+$ and $He^+$ ions induces alterations in microstructural integrity. Under irradiation conditions, NBG-17 graphite remained relatively stable even at high fluence levels, exhibiting microstructural resilience. Irradiation of IG-110 and NCBJ induced significant alterations in morphology, notably, most substantial in NCBJ graphite, confirming its sensitivity to ion impact.

*Declaration of competing interest*

*The authors declare that they have no known competing financial interests or personal relationships that could have appeared to influence the work reported in this paper.*

*CRediT authorship contribution statement*

M. Wilczopolska: Conceptualization, Methodology, Visualization, Formal analysis, Writing – original draft, Writing – review&editing; K. Suchorab: Conceptualization, Methodology, Visualization, Formal analysis, Writing – original draft, Writing – review&editing; M. Gawęda: Conceptualization, Visualization, Formal analysis, Writing – original draft, Writing – review&editing; M. Frelek-Kozak: Conceptualization; P. Ciepielewski: Conceptualization, Methodology; M. Brykała: Visualization; W. Chmurzyński: Visualization; I. Jóźwik: Conceptualization, Formal analysis, Writing – review&editing;

*Acknowledgements*

Research was partially funded through the European Union Horizon 2020 research and innovation program under Grant Agreement No. 857470 and from the European Regional Development Fund under the program of the Foundation for Polish Science International Research Agenda PLUS, grant No. MAB PLUS/2018/8, and the initiative of the Ministry of Science and Higher Education 'Support for the activities of Centers of Excellence established in Poland under the Horizon 2020 program' under agreement No. MEiN/2023/DIR/3795. The authors were supported by the Ministry of Education and Science founder of the „HTGR" project (agreement no. 1/HTGR/2021/14).